\documentclass[journal]{IEEEtran}

\usepackage[T1]{fontenc}
\usepackage[utf8]{inputenc}

\usepackage{cite}
\usepackage{graphicx}
\graphicspath{{assets/figures/}{assets/imgs/}}
\usepackage{array}
\usepackage{tabularx}
\newcolumntype{Y}{>{\raggedright\arraybackslash}X}
\usepackage{enumitem}

\usepackage{amsmath,amssymb}
\usepackage{siunitx}
\usepackage{booktabs}
\usepackage{multirow}
\usepackage{url}

\usepackage[font=footnotesize]{caption}
\DeclareCaptionType{listing}[List of listings][Listing]
\usepackage{stfloats}
\usepackage{placeins}
\usepackage{balance}

\usepackage{xcolor}
\usepackage{tikz}
\usetikzlibrary{arrows.meta,positioning,fit}

\usepackage{listings}
\lstdefinestyle{code}{
  basicstyle=\ttfamily\footnotesize,
  numbers=left, numberstyle=\tiny, stepnumber=1, numbersep=6pt,
  showstringspaces=false, keepspaces=true,
  breaklines=true, columns=fullflexible,
  frame=single, tabsize=2,
  xleftmargin=0pt, xrightmargin=0pt,
  aboveskip=4pt, belowskip=4pt
}
\lstdefinelanguage{json}{
  basicstyle=\ttfamily\footnotesize,
  showstringspaces=false,
  breaklines=true,
  breakatwhitespace=false,
  columns=fullflexible,
  literate=
   *{0}{{0}}1 {1}{{1}}1 {2}{{2}}1 {3}{{3}}1 {4}{{4}}1
    {5}{{5}}1 {6}{{6}}1 {7}{{7}}1 {8}{{8}}1 {9}{{9}}1
    {:}{{:}}1 {,}{{,}}1 {\{}{{\{}}1 {\}}{{\}}}1 {[}{{[}}1 {]}{{]}}1
}
\newcommand{\safeinclude}[2]{%
  \IfFileExists{#1}{%
    \includegraphics[width=#2]{#1}%
  }{%
    \fbox{%
      \parbox[c][2.5cm][c]{#2}{\centering\small File not found:\\\texttt{\detokenize{#1}}}%
    }%
  }%
}

\usepackage[flushleft]{threeparttable}

\usepackage[colorlinks=true,linkcolor=blue,citecolor=blue,urlcolor=blue]{hyperref}

\title{UPV\_RIR\_DB: A Structured Room Impulse Response Database with Hierarchical Metadata and Acoustic Indicators}

\author{\IEEEauthorblockN{Jes\'{u}s Garc\'{i}a-Gamborino$^{\dagger}$, Laura Fuster$^{\dagger}$, Daniel de la Prida$^{\diamond}$, Luis A. Azpicueta-Ruiz$^{\star}$, Gema Pi\~{n}ero$^{\dagger}$}\\
\IEEEauthorblockA{$^{\dagger}$ ITEAM, Universitat Polit\`{e}cnica de Val\`{e}ncia, Spain, jegarga3@upv.edu.es, \{lfuster,gpinyero\}@iteam.upv.es \\
$^{\diamond}$ Grupo de Investigaci\'{o}n en Ac\'{u}stica Arquitect\'{o}nica,  Universidad Polit\'{e}cnica de Madrid, Spain\\
$^{\star}$ Dep. Teor\'{i}a de la Se\~{n}al y Comunicaciones, Universidad Carlos III de Madrid, Spain}}
\begin{document}
\maketitle


\begin{abstract}
This paper presents UPV\_RIR\_DB, a structured database of measured room impulse responses (RIRs) designed to provide acoustic data with explicit spatial metadata and traceable acquisition parameters. The dataset currently contains 166 multichannel RIR files measured in three rooms of the Universitat Polit\`{e}cnica de Val\`{e}ncia (UPV).  Each multichannel RIR file contains impulse responses for multiple source–receiver pairs, with each pair covering a 25 cm$^2$ area — the typical size of a personal sound zone. Considering the number of sources and receiver channels associated with each microphone modality, the database contains a total of 18,976 single impulse responses. A hierarchical organization is adopted in which directory structure and metadata jointly describe the measurement context. Each room includes a metadata file containing acquisition parameters, hardware description, spatial coordinates of zones and microphones, and acoustic indicators such as reverberation time. A central index links each RIR file with its experimental context, ensuring traceability and enabling reproducible analysis. The resulting database provides a consistent framework for storing, inspecting, and reusing real RIR measurements while preserving compatibility with both MATLAB- and JSON-based workflows. The UPV\_RIR\_DB dataset is publicly available through the open repository Zenodo.
\end{abstract}



\section{Introduction}

Room Impulse Responses (RIRs) describe how sound propagates from a source to a receiver inside an enclosure. They contain direct sound, early reflections, and reverberant decay that characterize the acoustic behavior of a space. From measured RIRs, a variety of acoustic indicators can be derived, including reverberation time (T20/T30/T60), early decay time (EDT), clarity indices (C50/C80), definition (D50), and Speech Transmission Index (STI)~\cite{IEC60268-16}. These metrics are widely used in architectural acoustics, room analysis, auralization, and audio system evaluation~\cite{Kuttruff2016,ISO18233,ISO3382-1,Schroeder1965}.

The availability of publicly accessible RIR datasets has increased in recent years, supporting research in areas such as speech processing, acoustic modeling, and spatial audio. However, many existing collections exhibit limitations related to heterogeneous metadata, inconsistent naming conventions, or incomplete documentation of measurement conditions. In several cases, the lack of explicit spatial coordinates, calibration references, or acquisition parameters complicates reproducibility and cross-dataset comparisons.

These limitations motivate the development of structured datasets in which measurement context, spatial geometry, and acquisition parameters are clearly documented. In this work, we present UPV\_RIR\_DB, a multichannel RIR database measured in three rooms of the Universitat Polit\`{e}cnica de Val\`{e}ncia (UPV). The dataset is organized using a hierarchical directory structure combined with normalized metadata files that store the spatial descriptors and acquisition parameters required to interpret each measurement.

The goal of this dataset is to provide a reproducible and extensible repository of measured RIRs where each impulse response can be traced to its exact experimental configuration, including room condition, loudspeaker configuration, microphone type, spatial position, and calibration information.

The complete dataset is publicly available through the Zenodo research data repository~\cite{upv_rir_db_2026}. 

\subsection{Motivation and Objectives}

The design of the UPV\_RIR\_DB database follows three main objectives:

\begin{itemize}
\item \textbf{Structured organization.} Measurements are organized using a directory structure based on room configuration, loudspeaker configuration, and microphone modality.
\item \textbf{Hierarchical metadata.} Each room dataset includes a metadata file that stores spatial descriptors (zones, loudspeakers, microphones), acquisition parameters, and acoustic indicators.
\item \textbf{Traceability and reproducibility.} A central index links each stored RIR file with its measurement context, ensuring that every response can be traced back to its experimental conditions.
\end{itemize}

The resulting database provides a consistent framework for storing and analyzing RIR measurements while remaining compatible with existing MATLAB-based and JSON-based data processing workflows.

\subsection{Paper Organization}

The remainder of this article is organized as follows:

\begin{enumerate}[label=\roman*)]
\item Section~\ref{sec:related} reviews representative public RIR datasets and measurement standards.
\item Section~\ref{sec:acquisition} summarizes the rooms, equipment, and acquisition procedure used to collect them.
\item Section~\ref{sec:design} describes the database design, directory hierarchy, and metadata organization measurements.
\item Section~\ref{sec:concl} concludes the paper.
\end{enumerate}


\section{Public Datasets and Standards}
\label{sec:related}

Public datasets of Room Impulse Responses (RIRs) have become an essential resource for research in room acoustics, speech processing, spatial audio, and acoustic modeling. Several collections provide measured impulse responses under different spatial sampling strategies, microphone configurations, and acoustic environments.

However, despite their value, many datasets focus primarily on specific applications and often lack standardized metadata structures that clearly describe acquisition geometry, calibration references, or measurement conditions. In this section, we briefly review representative datasets and the measurement standards commonly adopted in room acoustics.

\subsection{Overview of public datasets}

Several publicly available datasets illustrate the diversity of acquisition approaches and application targets in RIR research.

\textbf{OpenAIR} provides a large collection of impulse responses primarily intended for auralization and audio production. The database includes measurements from a variety of spaces, such as cathedrals, halls, and studios, and offers both mono and binaural recordings. Although widely used in music technology, the dataset mainly provides descriptive documentation rather than structured metadata~\cite{OpenAIR}.

\textbf{BUT-ReverbDB} was designed for automatic speech recognition (ASR) research and contains real RIRs, background noise recordings, and retransmitted speech signals. Measurements were performed in several rooms using both mono and binaural microphones, and the dataset includes detailed acquisition documentation oriented toward speech processing experiments~\cite{BUTReverb}.

\textbf{MeshRIR} focuses on dense spatial sampling of impulse responses using two-dimensional and three-dimensional grids of measurement points. The dataset includes large collections of impulse responses captured with microphone arrays and is commonly used to evaluate sound field reconstruction and spatial interpolation techniques~\cite{MeshRIR}.
\begin{table*}[!t]
\renewcommand{\arraystretch}{1.1}
\caption{Overview of representative public RIR/BRIR datasets}
\label{tab:rir_overview}
\centering
\resizebox{\textwidth}{!}{%
\begin{tabular}{lcccccl}
\toprule
\textbf{Dataset} & \textbf{Year} & \textbf{\#RIRs} & \textbf{fs (Hz)} & \textbf{Microphones} & \textbf{License} & \textbf{Notes} \\
\midrule
OpenAIR~\cite{OpenAIR} & 2010-- & $>$500 & 44,100 & Mono / Binaural & CC BY & Auralization, audio design \\
BUT-ReverbDB~\cite{BUTReverb} & 2018--2019 & $\sim$10,000 & 16,000 / 48,000 & Mono / Binaural & Custom & ASR and speech processing \\
MeshRIR~\cite{MeshRIR} & 2021 & $\sim$200,000 & 48,000 & Arrays (4--16 ch) & CC BY & Dense spatial sampling \\
MOTUS~\cite{MOTUS2021I3DA} & 2021 & $\sim$13,000 & 48,000 & HOA & CC BY / Custom & Ambisonics measurements \\
Multizone RIR~\cite{Zhao2022MultizoneRIR} & 2022 & $\sim$260,000 & 48,000 & Arrays & CC BY & Multizone reproduction \\
\bottomrule
\end{tabular}
}
\end{table*}
\begin{table*}[!t]
\renewcommand{\arraystretch}{1.1}
\caption{Measurement methods and standards in representative RIR/BRIR datasets}
\label{tab:rir_methods}
\centering
\resizebox{0.8\textwidth}{!}{%
\begin{tabular}{lcccccc}
\toprule
\textbf{Dataset} & \textbf{Signal} & \textbf{Standard} & \textbf{Bands} & \textbf{Calibration} & \textbf{Extensible} & \textbf{Post-proc.} \\
\midrule
OpenAIR~\cite{OpenAIR} & Sweep & -- & Octave & Partial & No & -- \\
BUT-ReverbDB~\cite{BUTReverb} & Sweep / MLS & -- & -- & Partial & No & No \\
MeshRIR~\cite{MeshRIR} & Log Sweep & ISO 18233 & Octave & Yes & No & Scripts \\
MOTUS~\cite{MOTUS2021I3DA} & Log Sweep & -- & Octave & Yes & No & -- \\
Multizone RIR~\cite{Zhao2022MultizoneRIR} & Log Sweep & -- & Octave & Yes & No & -- \\
\bottomrule
\end{tabular}
}
\end{table*}
\textbf{MOTUS} provides a multichannel dataset recorded with higher-order ambisonics microphones under a large number of room configurations, including different furniture layouts. The dataset links acoustic measurements with synchronized three-dimensional models of the measured space, enabling realistic spatial audio rendering~\cite{MOTUS2021I3DA}.

\textbf{Multizone RIR} introduces a large-scale dataset measured with circular loudspeaker arrays and microphone arrays distributed across multiple listening zones. The collection contains hundreds of thousands of impulse responses intended for multizone sound field reproduction research~\cite{Zhao2022MultizoneRIR}.

Tables~\ref{tab:rir_overview} and~\ref{tab:rir_methods} summarize the main parameters of the five datasets described above. Although these datasets provide valuable measurement data, many lack a unified metadata model capable of describing acquisition parameters, spatial coordinates, calibration references, and measurement context in a consistent and reproducible manner.
\subsection{Measurement methods and standards}

Room acoustic measurements are typically performed following guidelines such as \textbf{ISO 18233}, which specifies measurement signals, acquisition procedures, and environmental conditions for impulse response measurements~\cite{ISO18233}. Acoustic indicators commonly derived from RIRs, such as reverberation time and clarity metrics, are defined in \textbf{ISO 3382-1}~\cite{ISO3382-1}.

Two excitation signals are widely used for RIR measurements:

\begin{itemize}
\item \textbf{Maximum Length Sequence (MLS)} signals, which allow fast measurements but are sensitive to nonlinear distortions.
\item \textbf{Sine sweeps}, particularly the \textbf{Exponential Sine Sweep (ESS)}, which provide improved signal-to-noise ratio and better separation of harmonic distortions~\cite{Farina2000,Mueller2001Sweeps}.
\end{itemize}

An ESS signal of duration $T$ ($0 \le t \le T$) sweeping from frequency $f_1$ to $f_2$ can be expressed as

\begin{equation}
x(t) =
\sin\!\left(
\frac{2\pi f_1 T}{\ln(f_2/f_1)}
\left(e^{\tfrac{t}{T}\ln(f_2/f_1)}-1\right)
\right).
\end{equation}

This exponential frequency mapping allocates more time to lower frequencies, improving measurement robustness and enabling separation of nonlinear distortion components during the deconvolution stage.

Calibration procedures and documentation of measurement conditions remain essential for reproducibility. Recording reference levels, timestamps, and equipment configurations ensures traceability across measurement sessions and datasets~\cite{ISO18233,ISO3382-1}.

\subsection{Identified gaps and adopted criteria}

A comparison of existing datasets reveals several recurring limitations, although not all datasets present all of these issues simultaneously:

\begin{itemize}
\item Metadata is often incomplete or described only narratively.
\item Spatial coordinates of sources and receivers are not always available.
\item Calibration references and acquisition parameters are sometimes missing.
\item Directory structures and naming conventions are heterogeneous across datasets.
\end{itemize}

While some datasets provide highly detailed documentation in specific aspects, a consistent balance between clarity, simplicity, reproducibility, scalability, and completeness of information is not always achieved.

To address these issues, the design of UPV\_RIR\_DB adopts the following principles:

\begin{itemize}
\item A hierarchical directory structure.
\item A normalized metadata model describing spatial geometry, acquisition parameters, and hardware configuration.
\item Spatial coordinates for zones, loudspeakers, and microphones.
\item A central index linking each RIR file with its experimental context.
\end{itemize}

These criteria aim to ensure that each impulse response in the dataset can be interpreted and reused without ambiguity.


\section{RIR acquisition and experimental design}
\label{sec:acquisition}

This section describes the experimental system used to acquire the impulse responses included in UPV\_RIR\_DB, together with its room-dependent deployment.


\subsection{Measurement equipment}

The impulse responses were recorded using a multichannel acquisition system composed of audio interfaces, a loudspeaker array, and several microphone modalities. The specific configuration used in each campaign depends on the room and the associated experimental design. This subsection summarizes the hardware characteristics required to understand the acquisition process, whereas detailed device models and identifiers are provided in Appendix~\ref{app:hardware}.

\subsubsection*{Audio interfaces}

Recordings were carried out using multichannel audio interfaces with 16 input channels and 8 output channels. Depending on the room and the number of channels required by the acquisition modality, either a single interface or several synchronized interfaces were employed.

\subsubsection*{Loudspeaker array}

The excitation system consists of a horizontal linear array of 8 active two-way loudspeakers. The enclosures are placed contiguously, without physical spacing between cabinets, although the effective acoustic spacing is determined by the distance between the acoustic centers of the transducers, 0.19\,m. The exact loudspeaker coordinates are stored in the metadata of each room and configuration.

\begin{table*}[t]
\footnotesize
\centering
\caption{Measurement hardware and acquisition configurations by room. The metadata of each room explicitly documents the hardware configuration, loudspeaker layouts, and microphone modalities associated with each measurement campaign.}
\label{tab:hardware_capture_summary}
\renewcommand{\arraystretch}{1.1}
\setlength{\tabcolsep}{4pt}
\begin{threeparttable}
\begin{tabularx}{\textwidth}{@{}l Y Y Y Y@{}}
\toprule
\textbf{Room} & \textbf{Audio interface(s)} & \textbf{Loudspeakers} & \textbf{Microphone modalities} & \textbf{Notes} \\
\midrule
4P13 & 16×8 interface & 8-source active 2-way array (SC1) & AR, DH & Includes four room configurations (RC1--RC4). \\
\midrule
4D26 & 16×8 interface & 8-source active 2-way array (SC1) & AR, DH, BI & Includes trajectory-based measurements. \\
\midrule
8G4C & Dual 16×8 interfaces & 8-source active 2-way array (SC1, SC2) & AR (dual array) & Two synchronized 16-element arrays used in parallel; SC1 and SC2 correspond to two placements of the same loudspeaker array. \\
\bottomrule
\end{tabularx}
\end{threeparttable}
\end{table*}

\subsubsection*{Microphone modalities}

The database employs three main microphone modalities: \textbf{AR}, \textbf{DH}, and \textbf{BI}.

The \textbf{AR} modality corresponds to a regular planar array of 16 microphones arranged in a \(4 \times 4\) matrix, with an inter-element spacing of 0.07\,m along both the \(x\)- and \(y\)-axes. The nominal microphone diameter is 12.7\,mm. This internal geometry remains constant throughout the corpus.

The \textbf{DH} modality corresponds to a \textit{dummy head}, used as a rigid binaural receiver with a fixed geometry.

The \textbf{BI} modality corresponds to a binaural headset with two miniature microphones placed at aural positions. Its internal geometry remains fixed, although this modality incorporates additional experimental conditions and local per-measurement documentation, described later.

\begin{figure*}[!t]
\centering
\begin{minipage}[t]{0.31\textwidth}
\centering
\includegraphics[width=\linewidth]{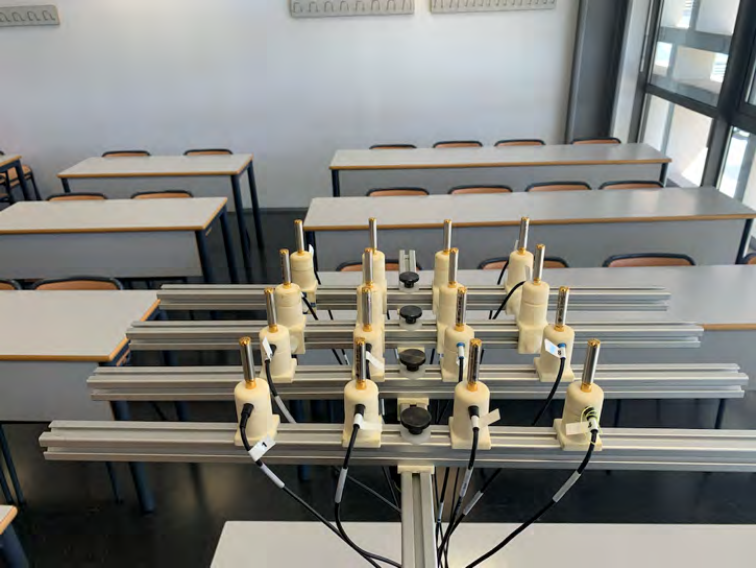}\\[-1pt]
\footnotesize (a) 
\end{minipage}
\hfill
\begin{minipage}[t]{0.31\textwidth}
\centering
\includegraphics[width=\linewidth]{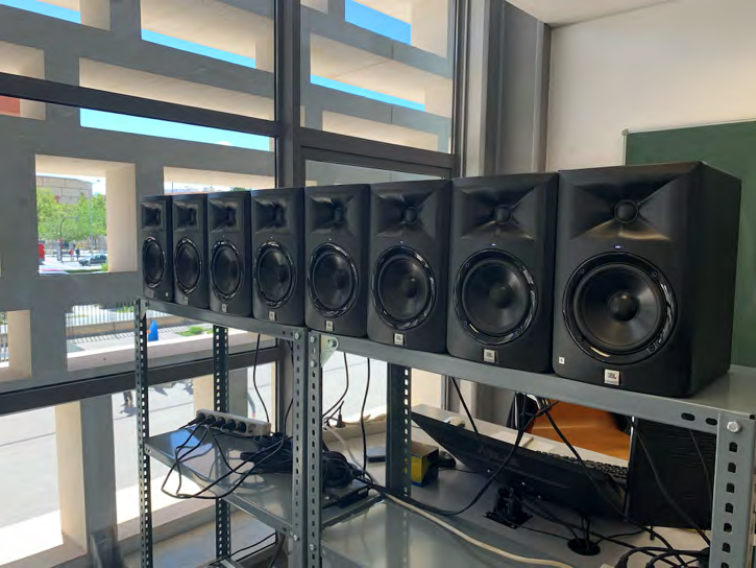}\\[-1pt]
\footnotesize (b) 
\end{minipage}
\hfill
\begin{minipage}[t]{0.31\textwidth}
\centering
\includegraphics[width=\linewidth]{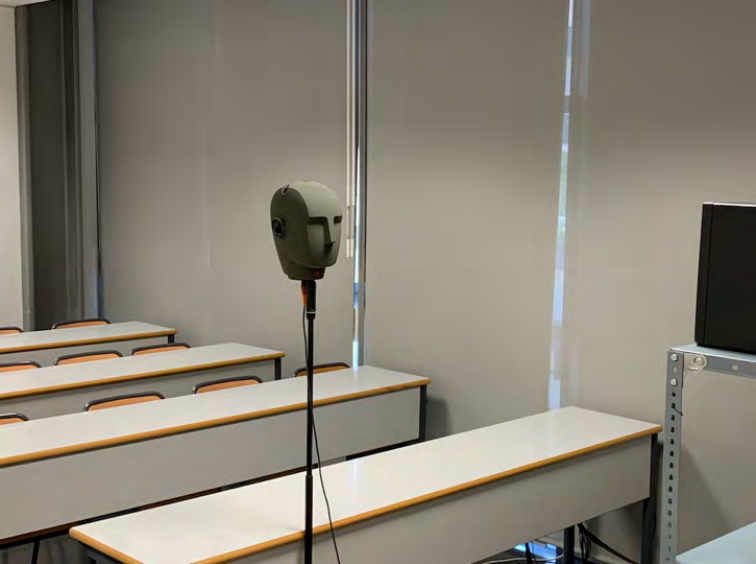}\\[-1pt]
\footnotesize (c) 
\end{minipage}
\caption{Measurement hardware used in the acquisition campaigns: (a) 4x4 array microphone, (b) 8-source loudspeaker array, and (c) Neumann KU100 Dummy Head.}
\label{fig:hardware_composite}
\end{figure*}

Table~\ref{tab:hardware_capture_summary} summarizes the main hardware and acquisition configurations used in the current release of the database, whereas Fig.~\ref{fig:hardware_composite} shows the main hardware equipment used in the acquisition campaign.

\subsection{Measured rooms and acquisition methodology}

\begin{figure*}[!t]
\centering
\begin{minipage}[t]{0.31\textwidth}
\centering
\includegraphics[width=\linewidth]{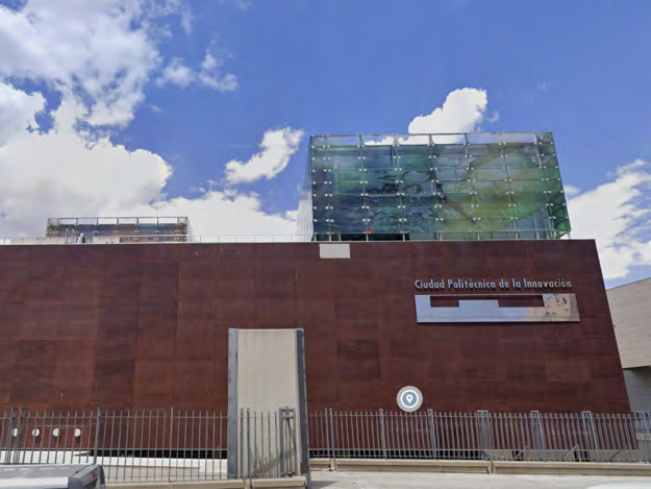}\\[-1pt]
\footnotesize (a) 
\end{minipage}
\hfill
\begin{minipage}[t]{0.31\textwidth}
\centering
\includegraphics[width=\linewidth]{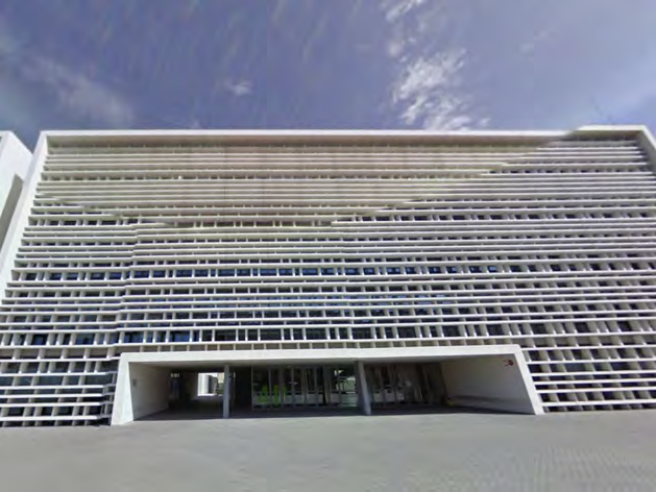}\\[-1pt]
\footnotesize (b) 
\end{minipage}
\hfill
\begin{minipage}[t]{0.31\textwidth}
\centering
\includegraphics[width=\linewidth]{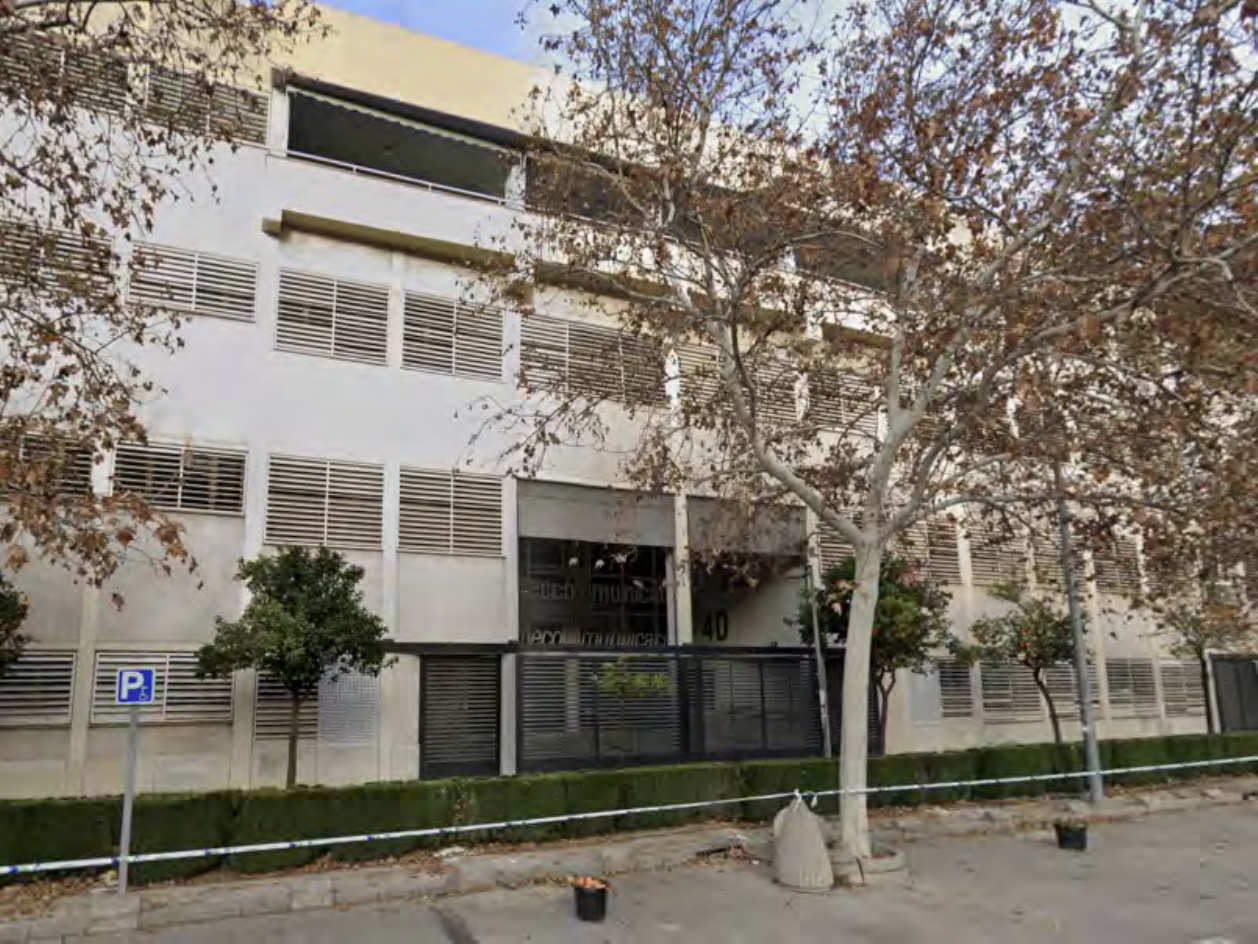}\\[-1pt]
\footnotesize (c) 
\end{minipage}
\caption{Buildings housing the rooms used for the database. In parenthesis the acronym of the corresponding room: (a) iTEAM (8G4C), (b) ETSIT Building 4P (4P13), and (c) ETSIT Building 4D (4D26).}
\label{fig:room_buildings}
\end{figure*}

The measurement design in each room was conditioned by three main factors: (i) the available experimental system, (ii) the geometric and functional characteristics of the enclosure, and (iii) the specific objectives of each campaign. As a result, although the internal geometry of the microphone modalities and the loudspeaker array remains stable, the selection of zones, points, extensions, and trajectories varies from room to room. Fig.~\ref{fig:room_buildings} shows the buildings located at the UPV campus housing the rooms used for the database.

\paragraph{Room\_iTEAM\_8G4C\_mat}
This room corresponds to a space at the \textit{Institute of the Telecommunications and Multimedia Applications} (iTEAM), located within the East part of the UPV campus\footnote{Google maps: \url{https://maps.app.goo.gl/y38bbCxJxkZtUnVm8}}. Its dimensions are 3.58\,m \(\times\) 7.78\,m \(\times\) 2.80\,m.
\begin{figure}[!t]
\centering
\includegraphics[width=\columnwidth]{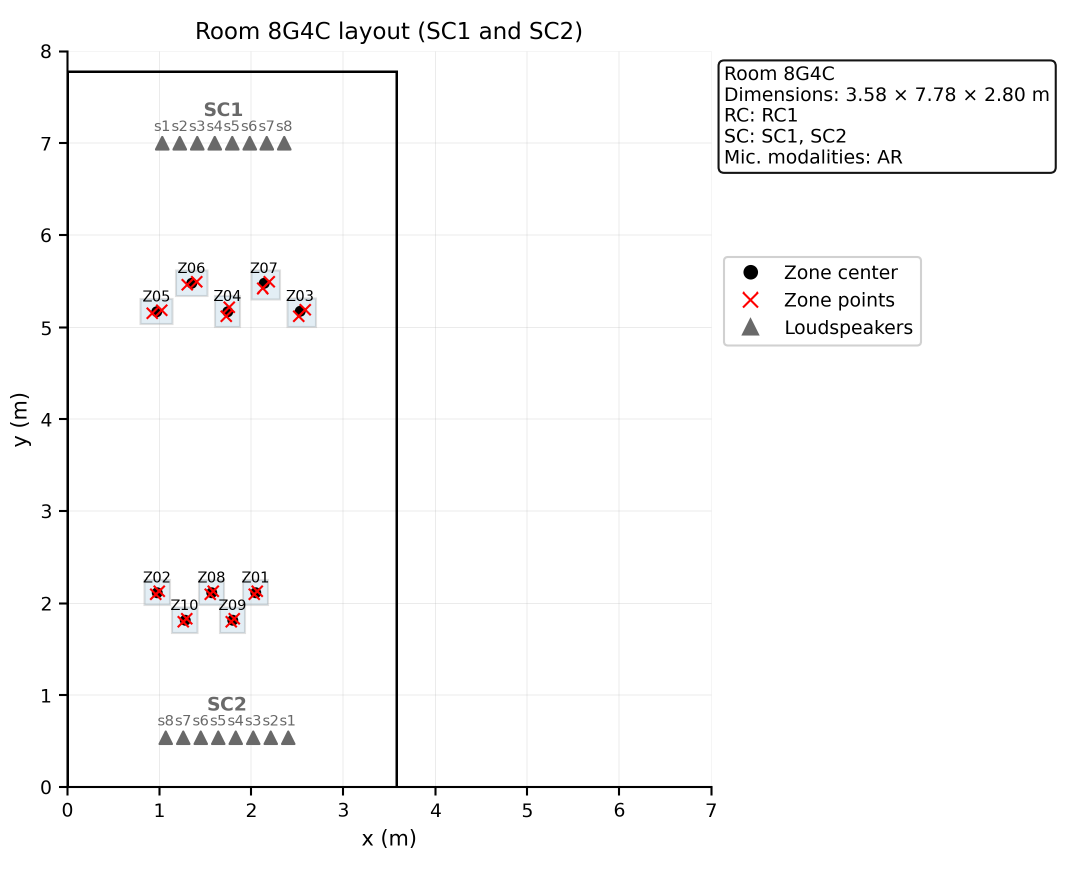}
\caption{Measurement layout used in Room 8G4C. Zones \texttt{Z03}--\texttt{Z07} are associated with \texttt{SC1}, whereas zones \texttt{Z01}, \texttt{Z02}, and \texttt{Z08}--\texttt{Z10} are associated with \texttt{SC2}.}
\label{fig:8g4c_layout}
\end{figure}

\begin{figure}[!t]
\centering
\includegraphics[width=\columnwidth]{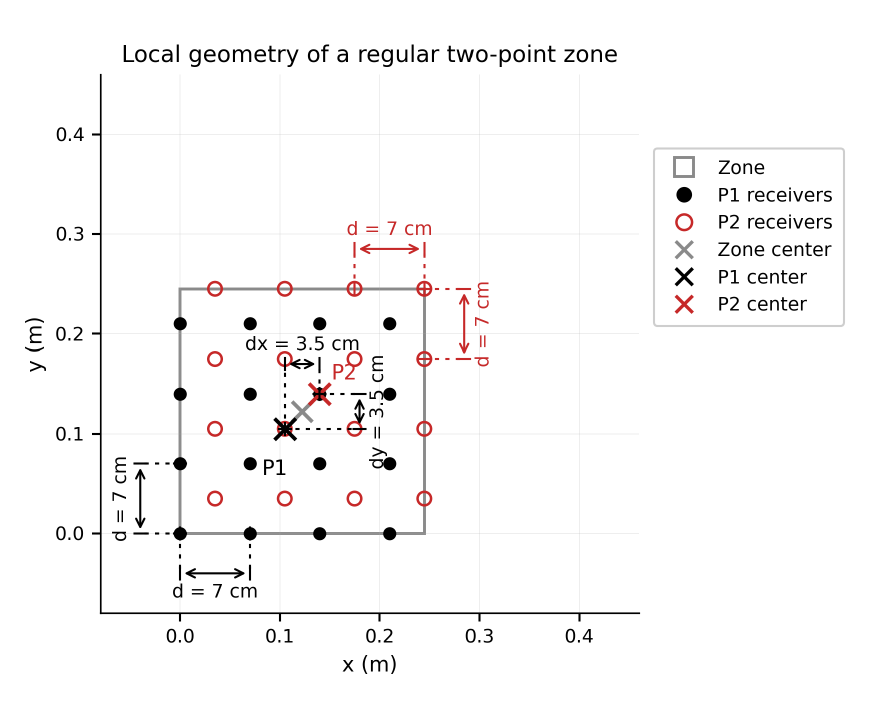}
\caption{Local measurement scheme for a regular two-point zone. The AR modality uses a \(4 \times 4\) grid with 0.07\,m spacing, and the offset between \texttt{P01} and \texttt{P02} is 0.035\,m in both \(x\) and \(y\).}
\label{fig:zone_grid_detail}
\end{figure}

\begin{figure}[!t]
\centering
\includegraphics[width=\columnwidth]{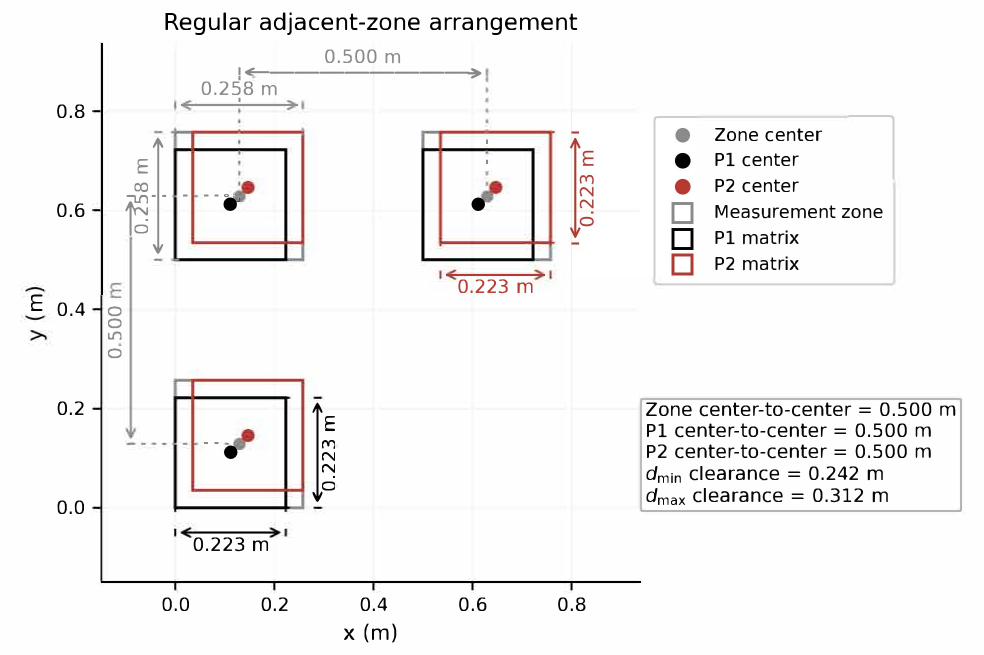}
\caption{Regular adjacent-zone extension used in the basic measurement layout of Rooms 4P13 and 4D26. Adjacent zone centers are spaced by 0.50\,m in both \(x\) and \(y\).}
\label{fig:adjacent_zone_extension}
\end{figure}

\begin{figure}[!t]
\centering
\includegraphics[width=\columnwidth]{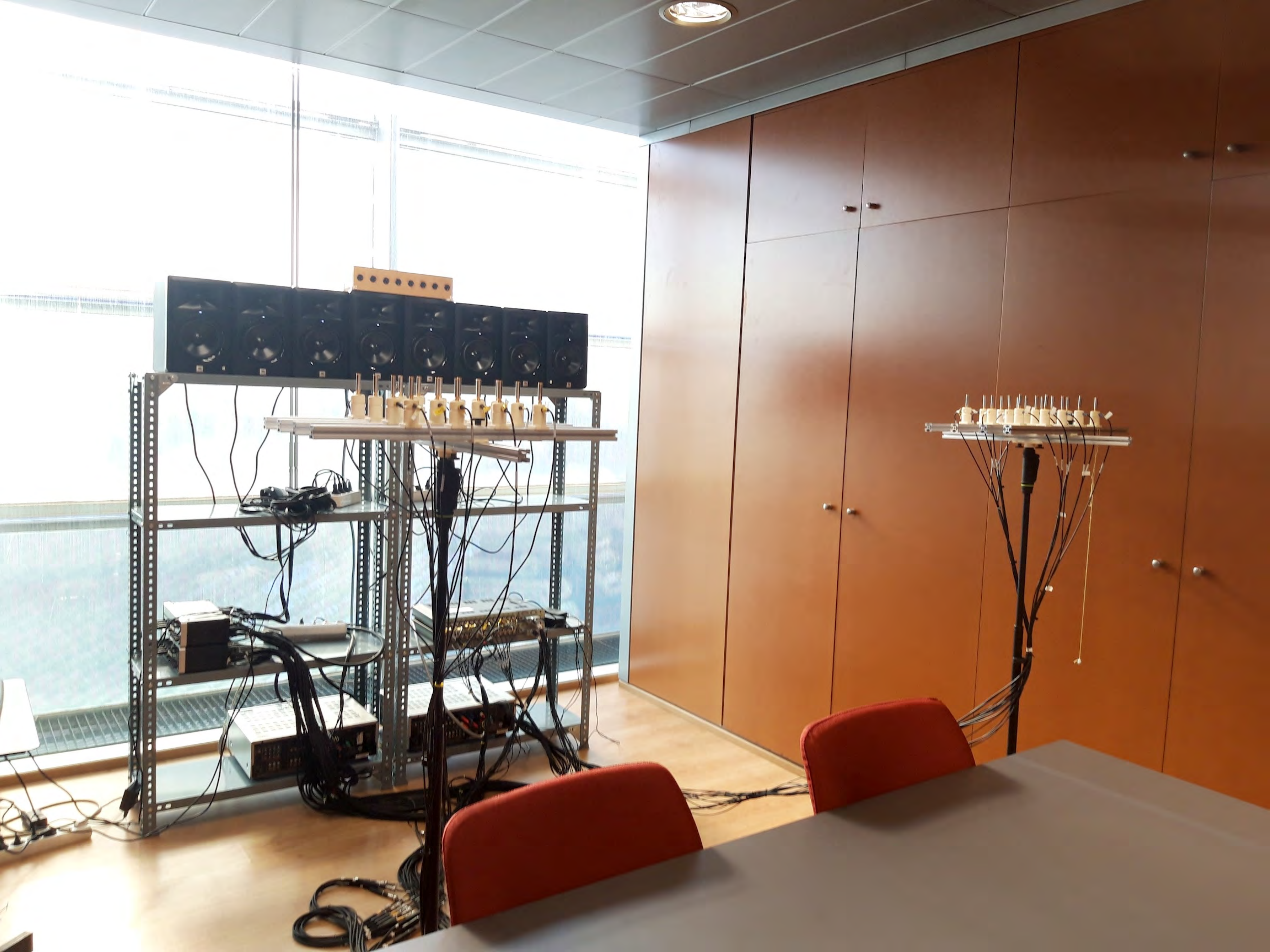}
\caption{General view of the measurement setup in Room 8G4C.}
\label{fig:8G4C_photo}
\end{figure}

Acquisition in this room was carried out exclusively with the \textbf{AR} modality, using two 16-element microphone arrays. The dataset comprises 10 measurement zones, with two sets of $8 \times 16$ measured RIRs per zone, distributed according to the layout shown in Fig.~\ref{fig:8g4c_layout}. For each zone, the microphone array was placed at two different positions P1 and P2 according to the schemes shown in Fig.~\ref{fig:zone_grid_detail} and.~\ref{fig:adjacent_zone_extension}. The dataset includes two loudspeaker configurations, \texttt{SC1} and \texttt{SC2}, which correspond to two placements of the same 8-source loudspeaker array located on opposite sides of the room and oriented toward the central measurement region. Zones \texttt{Z03}--\texttt{Z07} are associated with \texttt{SC1}, whereas zones \texttt{Z01}, \texttt{Z02}, and \texttt{Z08}--\texttt{Z10} are associated with \texttt{SC2}. The loudspeaker height is 1.565\,m, whereas the microphone-array height is 1.50\,m. Fig.~\ref{fig:8g4c_layout} shows the Measurement layout used in Room 8G4C, while Fig.~\ref{fig:8G4C_photo} shows a picture of the equipment inside the room. 

At corpus level, this room includes 10 multichannel AR ($8 \times 32$) RIRs, corresponding to a total of $2560$ individual impulse responses.

\begin{figure}[!t]
\centering
\includegraphics[width=\columnwidth]{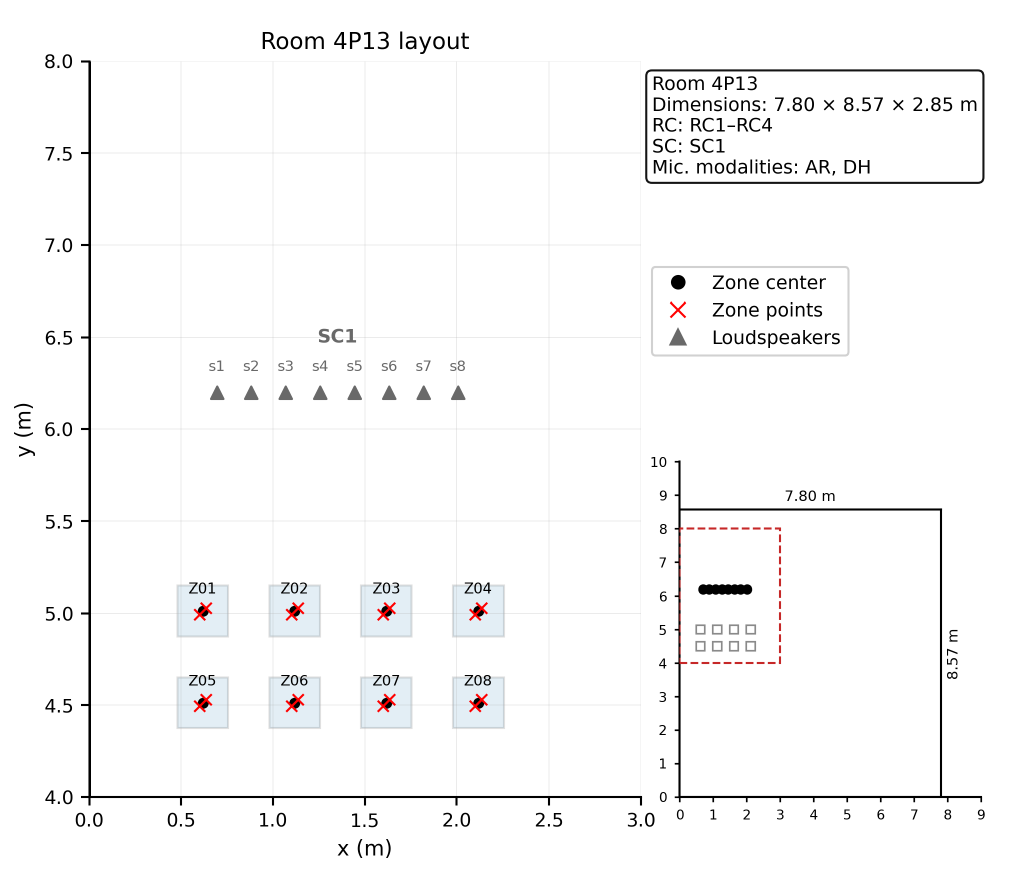}
\caption{Measurement layout of Room 4P13. The regular 8-zone organization is preserved across the four room configurations.}
\label{fig:4p13_layout}
\end{figure}

\begin{figure}[!t]
\centering
\includegraphics[angle=-90,width=\columnwidth]{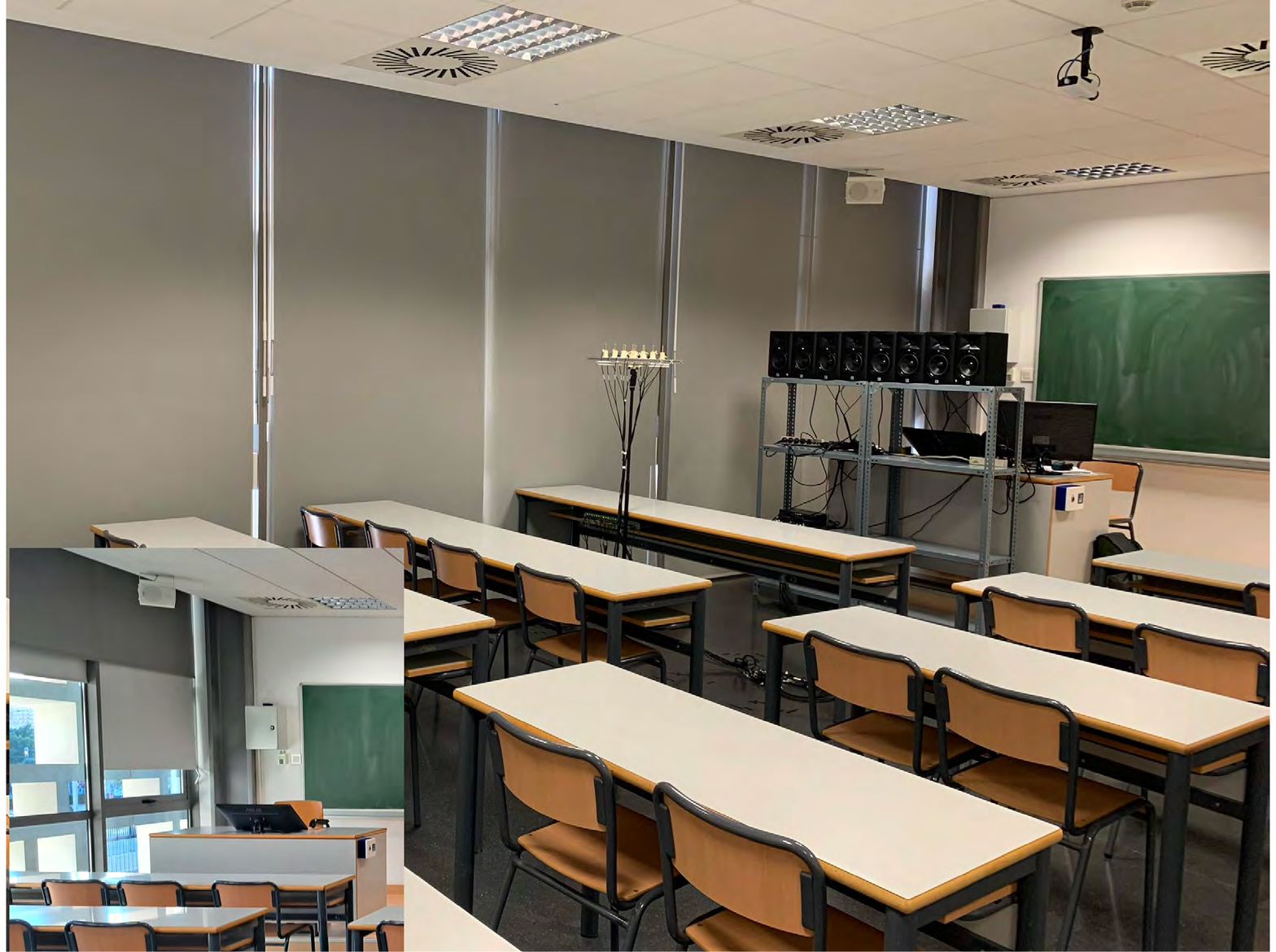}
\caption{General view of the measurement setup in Room 4P13. A detail of the window, the reflective surface and the curtain of the room can be seen at the bottom left.}
\label{fig:4p13_photo}
\end{figure}

\paragraph{Room\_Teleco\_4P13\_mat}
This room corresponds to classroom 13 on the first floor of Building 4P of the \textit{Escuela T\'{e}cnica Superior de Ingenieros de Telecomunicai\'{o}n} (ETSIT) located within the center part of the UPV campus\footnote{Google maps: \url{https://maps.app.goo.gl/1CdtAxxoAVPWwn216}}. Its dimensions are 7.80\,m \(\times\) 8.57\,m \(\times\) 2.85\,m. Fig.~\ref{fig:4p13_layout} shows the measurement layout for room 4P13. A general view of the room and the measurement equipment is shown in Fig.~\ref{fig:4p13_photo}.

The dataset associated with this room includes four room configurations (\texttt{RC1}--\texttt{RC4}), defined according to two factors: the state of the glazed surface (window) and the presence or absence of very nearby reflective elements. A detail of the window, the reflective surface and the curtain of the room can be seen at the bottom left of Fig.~\ref{fig:4p13_photo}. Specifically:
\begin{itemize}
\item \texttt{RC1}: exposed or uncovered window area, without nearby reflective surface;
\item \texttt{RC2}: exposed or uncovered window area, with nearby reflective surface;
\item \texttt{RC3}: window area covered by curtain, without nearby reflective surface;
\item \texttt{RC4}: window area covered by curtain, with nearby reflective surface.
\end{itemize}
These four configurations represent different acoustic conditions within the same physical enclosure.

The room impulse responses were acquired under a single loudspeaker configuration (\texttt{SC1}) and two receiver modalities, \textbf{AR} and \textbf{DH}. Its spatial organization is based on a regular 8-zone arrangement, as shown in Fig.~\ref{fig:4p13_layout}, with two measurement points per zone (as shown in Fig.~\ref{fig:zone_grid_detail}), consistently reused across the four room configurations. In this regular scheme, points \texttt{P01} and \texttt{P02} are displaced by 0.035\,m along both \(x\) and \(y\). For the \textbf{AR} modality, each point is materialized by a local \(4 \times 4\) grid with 0.07\,m spacing between adjacent receivers along both axes. At the inter-zone level, adjacent zone centers are separated by 0.50\,m in both \(x\) and \(y\), defining the regular spatial base later reused in the 4D26 room extension. In this room, both loudspeakers and microphone systems were positioned at 1.65\,m. 

Measurements in this room were acquired using the \textbf{AR} and \textbf{DH} modalities under the same loudspeaker configuration. In total, the room includes 32 multichannel AR ($8 \times 32$)  RIRs and 32 multichannel DH ($8 \times 4$) RIRs, corresponding to 8192 and 1024 individual impulse responses, respectively.

\begin{figure}[!t]
\centering
\includegraphics[width=\columnwidth]{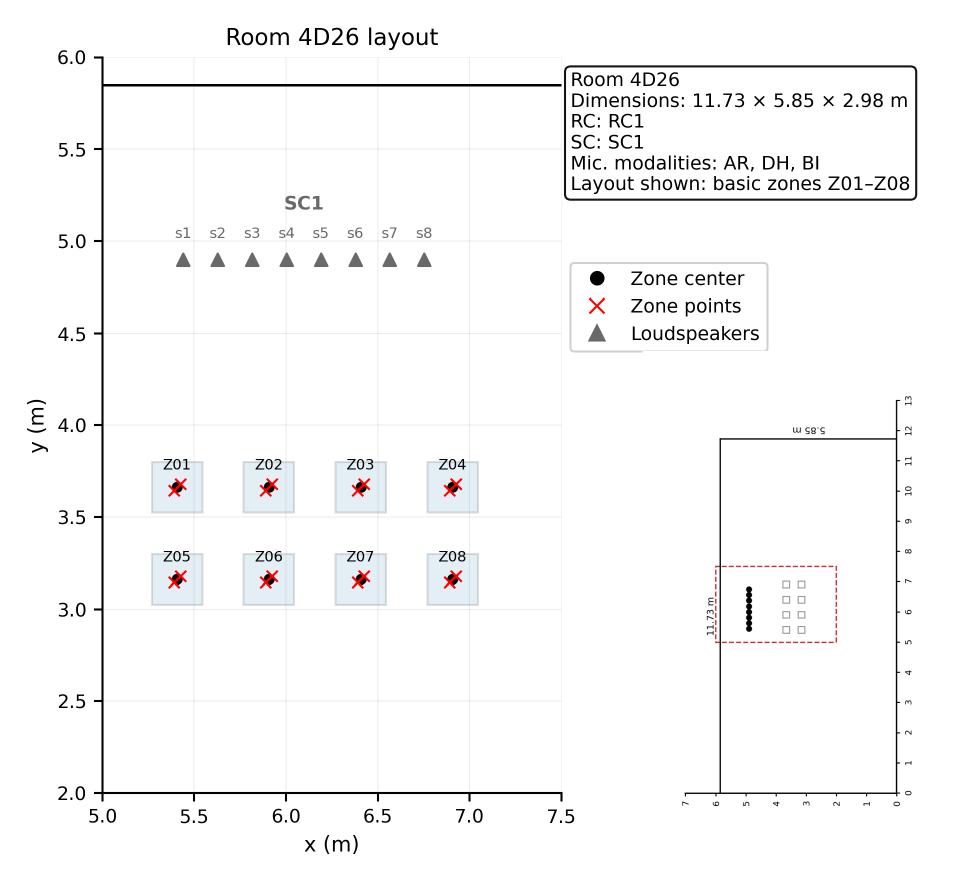}
\caption{Basic 8-zone layout used as the regular spatial base in Room 4D26.}
\label{fig:4d26_basic_layout}
\end{figure}

\begin{figure}[!t]
\centering
\includegraphics[width=\columnwidth]{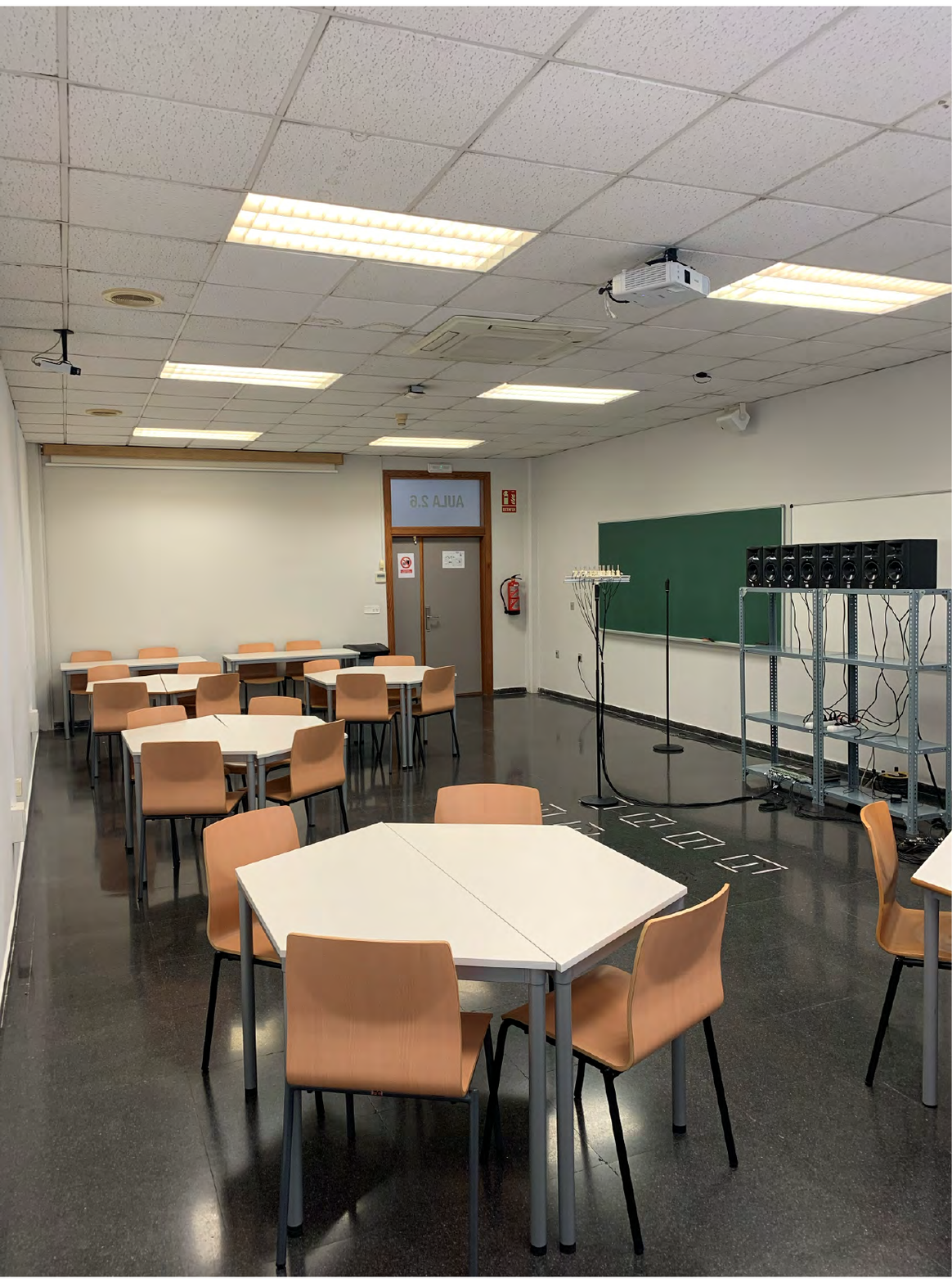}
\caption{General view of the measurement setup in Room 4D26.}
\label{fig:4d26_photo}
\end{figure}

\paragraph{Room\_Teleco\_4D26\_mat}
This room corresponds to classroom 26 on the second floor of Building 4D of the ETSIT located within the center part of the UPV campus\footnote{Google maps: \url{https://maps.app.goo.gl/RnFJWwc27Kmvnh9G8}}. Its dimensions are 11.73\,m \(\times\) 5.85\,m \(\times\) 2.98\,m. The basic layout of the loudspeaker array and the measured zones is shown in Fig.~\ref{fig:4d26_basic_layout}, while a general view of the room with the measurement system is shown in Fig.~\ref{fig:4d26_photo}.

Their room impulse responses were acquired under a single normalized room configuration (\texttt{RC1}) and a single loudspeaker configuration (\texttt{SC1}). In the dataset convention, \texttt{RC1} corresponds to an exposed or uncovered window condition without a nearby reflective surface.

The spatial organization of this room starts from a basic regular arrangement formed by zones \texttt{Z01}--\texttt{Z08}, following the same general base criterion used in 4P13. In this base structure, points \texttt{P01} and \texttt{P02} are displaced by 0.035\,m along both \(x\) and \(y\), and the center-to-center spacing between adjacent basic zones is 0.50\,m along both axes, as illustrated in Fig.~\ref{fig:zone_grid_detail} and~\ref{fig:adjacent_zone_extension}. For the \textbf{AR} and \textbf{DH} modalities, both loudspeakers and microphone systems were positioned at 1.65\,m.

From this initial structure, the zone catalog is extended with additional positions to support the trajectory-based acquisition design and PSZ-oriented layouts. In particular, zones \texttt{Z09}--\texttt{Z19} extend the spatial coverage to define diagonal trajectories, whereas zones \texttt{Z20}--\texttt{Z25} are introduced as frontal zones specific to the \textbf{BI} modality. As a result, zones \texttt{Z01}--\texttt{Z19} contain two measurement points per zone, while zones \texttt{Z20}--\texttt{Z25} contain a single point.

The metadata of 4D26 define six trajectories, \texttt{T01}--\texttt{T06}: trajectories \texttt{T01}--\texttt{T04} are diagonal, whereas \texttt{T05} and \texttt{T06} are perpendicular with respect to the line of the loudspeaker array. Based on the zone-center coordinates, the center-to-center distances between consecutive zones depend on each trajectory:
\begin{itemize}
\item \texttt{T01}: 0.318\,m, 0.371\,m, 0.336\,m, and 0.371\,m;
\item \texttt{T02}: 0.361\,m, 0.362\,m, 0.371\,m, 0.336\,m, and 0.371\,m;
\item \texttt{T03}: 0.361\,m, 0.347\,m, 0.358\,m, 0.351\,m, and 0.358\,m;
\item \texttt{T04}: 0.555\,m, 0.557\,m, and 0.562\,m;
\item \texttt{T05} and \texttt{T06}: 0.500\,m between consecutive zone centers.
\end{itemize}

The diagonal trajectories \texttt{T01}--\texttt{T04} are organized as two oblique families of opposite geometric sign over the extended zone set, whereas \texttt{T05} and \texttt{T06} define two frontal BI-only trajectories. In the metadata, \texttt{T05} and \texttt{T06} are described as frontal trajectories with \texttt{alignment = perpendicular} and \texttt{axis\_reference = y}. Their associated zones, \texttt{Z20}--\texttt{Z25}, form two frontal columns with fixed \(x\)-coordinates and three equally spaced \(y\)-positions, referenced to the projection of the loudspeaker array within the room coordinate system.

\begin{figure*}[!t]
\centering
\includegraphics[width=0.9\textwidth]{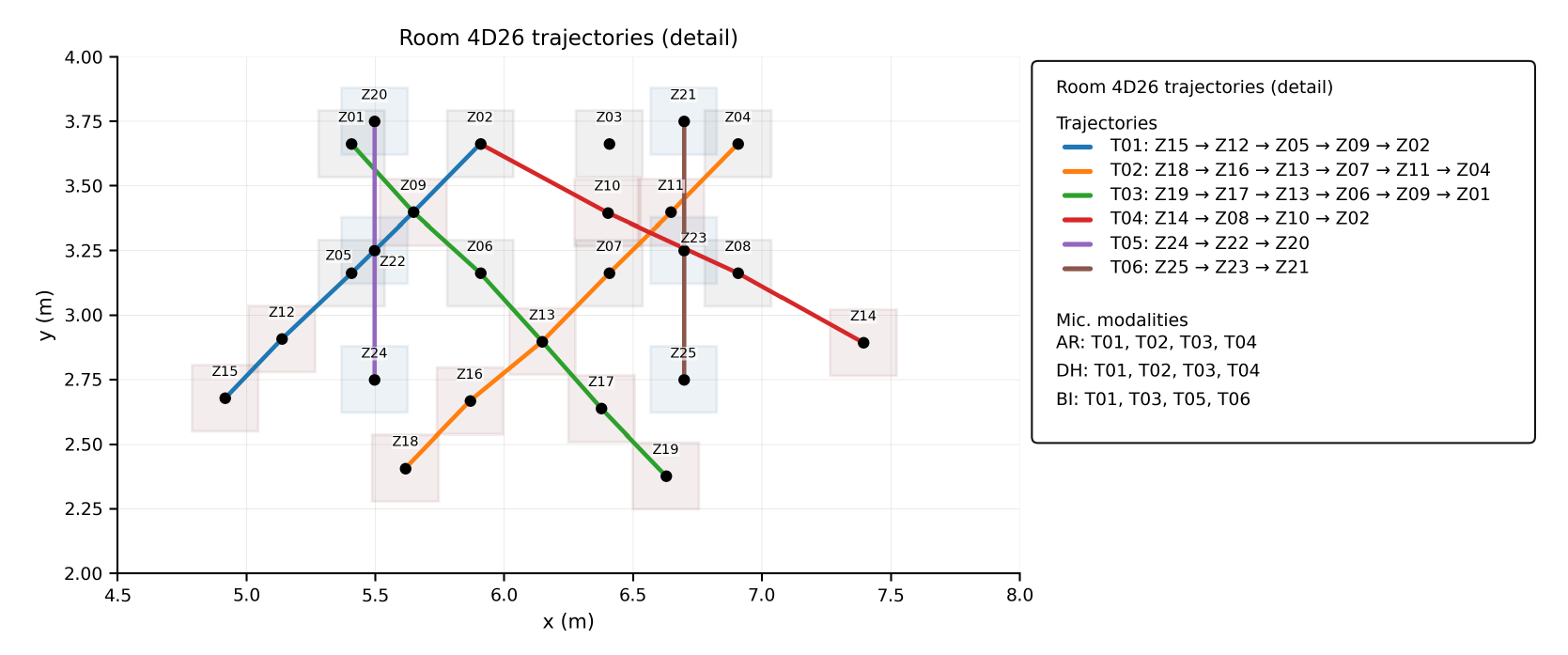}
\caption{Trajectory layout in Room 4D26. Diagonal trajectories are defined over the extended zone set, whereas frontal BI trajectories are organized separately.}
\label{fig:4d26_trajectories_detail}
\end{figure*}

The exact definition of each trajectory, including the ordered sequence of zones and its association with the available modalities, is stored in the metadata and is shown in Fig.~\ref{fig:4d26_trajectories_detail}. Table~\ref{tab:traj_4d26} summarizes the trajectories defined for room \texttt{4D26} and the microphone modalities with which they are used.

\begin{table}[!t]
\centering
\caption{Trajectory definitions in Room 4D26 according to the metadata}
\label{tab:traj_4d26}
\renewcommand{\arraystretch}{1.1}
\begin{tabular}{cll}
\toprule
Trajectory & Zones & Mic. types \\
\midrule
T01 & Z15, Z12, Z05, Z09, Z02 & AR, DH, BI \\
T02 & Z18, Z16, Z13, Z07, Z11, Z04 & AR, DH \\
T03 & Z19, Z17, Z13, Z06, Z09, Z01 & AR, DH, BI \\
T04 & Z14, Z08, Z10, Z02 & AR, DH \\
T05 & Z20, Z22, Z24 & BI \\
T06 & Z21, Z23, Z25 & BI \\
\bottomrule
\end{tabular}
\end{table}

Therefore, the spatial placement of zones in 4D26 is not globally uniform, but follows trajectory-dependent geometric relations. The resulting arrangement combines a regular 8-zone base, an extended diagonal zone set, and two frontal BI-only trajectories referenced to the loudspeaker-array geometry.


\begin{figure}[!t]
\centering
\includegraphics[angle=-90,width=\columnwidth]{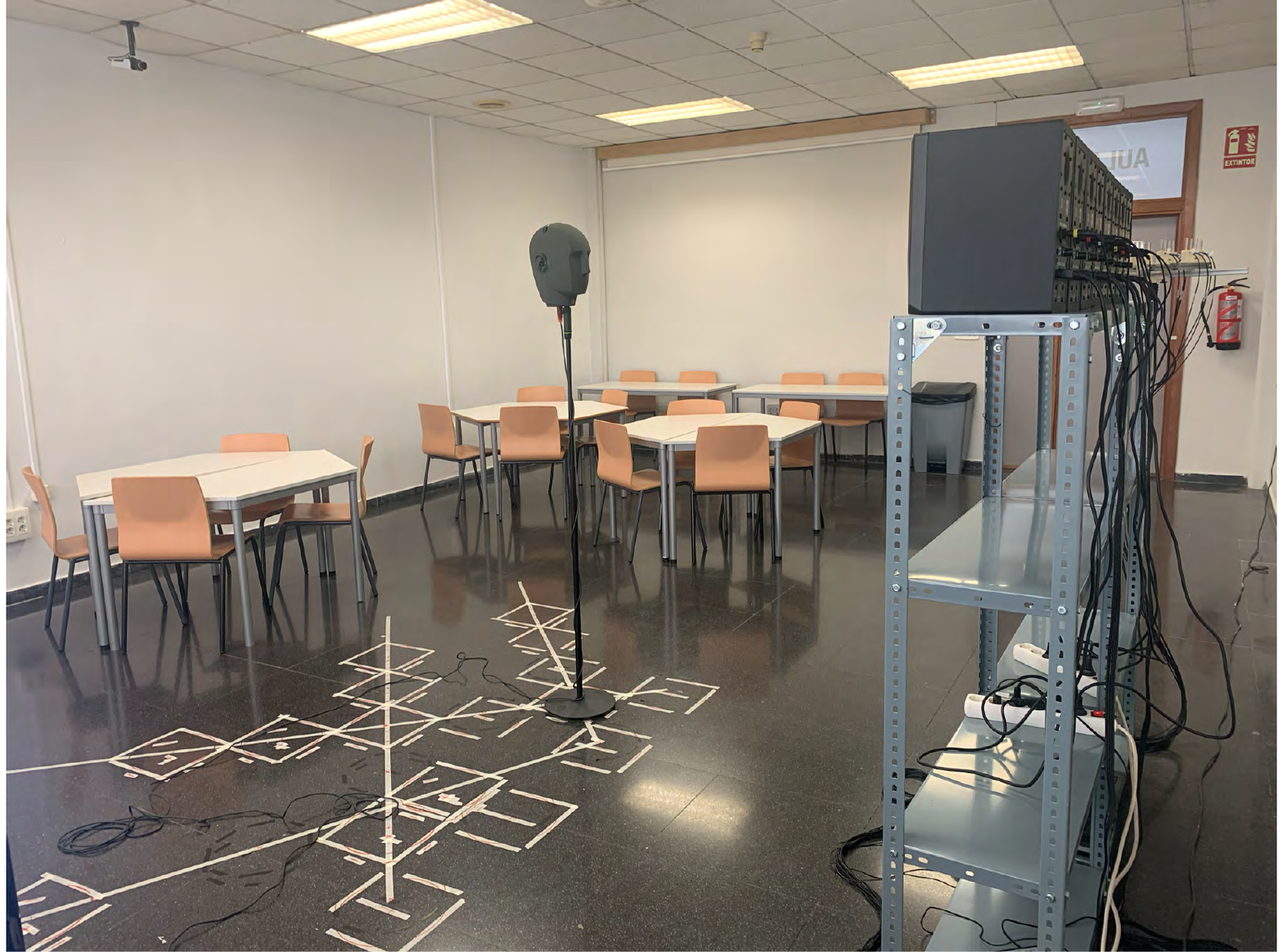}
\caption{Example of the trajectory-based deployment in Room 4D26, showing the DH configuration and the floor marking used for the measurement layout.}
\label{fig:4d26_dh_traj_photo}
\end{figure}

\begin{figure}[!t]
\centering
\includegraphics[width=\columnwidth]{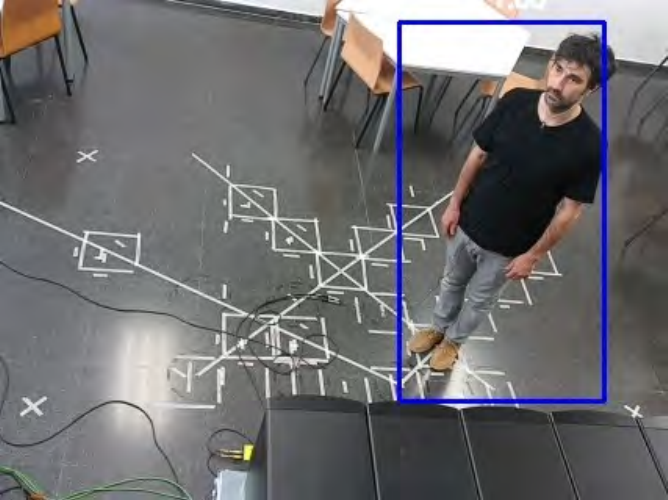}
\caption{Example of local per-RIR metadata associated with a BI measurement in Room 4D26. The snapshot provides visual traceability of the tracked subject position for a binaural acquisition case.}
\label{fig:bi_local_metadata_example}
\end{figure}

Room 4D26 is also the only room in the dataset in which the \textbf{BI} modality was employed. In this case, the measurements introduce additional orientation and height conditions. Receiver orientation is encoded as \textbf{C}, \textbf{L}, and \textbf{R}, whereas the two effective receiver-height conditions correspond to \textbf{F} (1.55\,m) and \textbf{M} (1.70\,m). The receiver-heights correspond to the distance from the ears of the person wearing the binaural microphones to the floor (male (M) or female (F)). In addition, each BI measurement includes local per-RIR documentation based on a captured image at the time of acquisition and a JSON file containing the corresponding spatial information. Fig.~\ref{fig:4d26_dh_traj_photo} shows a picture of the trajectory-based deployment in Room 4D26 when the dummy head was used, whereas Fig.~\ref{fig:bi_local_metadata_example} shows an example of the pictures included as metadata in the UPV\_RIR\_DB for each zone, receiver orientation, and receiver-height.

For BI measurements, the frontal condition corresponds to the receiver facing the loudspeaker array, whereas the lateral conditions correspond to path-oriented oblique cases defined relative to the trajectory geometry.

In total, this room includes 19 multichannel AR ($8 \times 32$) RIRs, 19 multichannel DH ($8 \times 4$) RIRs, and 54 multichannel BI ($8 \times 4$) RIRs, corresponding to 4864, 608, and 1728 individual impulse responses, respectively.


\subsection{Measurement conditions}

The \textbf{AR} and \textbf{DH} modalities were acquired under a fixed frontal orientation with respect to the loudspeaker array, ensuring consistent conditions across zones and configurations.

The \textbf{BI} modality incorporates additional receiver-orientation and height conditions, together with local per-RIR documentation intended to preserve the effective binaural acquisition context. These conditions are encoded in the corresponding metadata and file naming structure.


\subsection{Acquisition protocol}

The acquisition procedure was globally consistent across all rooms, with minor variations due to the available hardware and to the specific objectives of each campaign.

The impulse responses were acquired using sweep-based excitation signals. In this dataset, the classical exponential sweep principle described in the literature is implemented through the ESM configuration used in Rooms 4P13 and 4D26, whereas Room 8G4C uses the MESM configuration, corresponding to a multiple concatenated variant designed for synchronized dual-interface acquisition and simultaneous 32-channel capture.

The general acquisition protocol followed the steps summarized below:
\begin{enumerate}
\item \textbf{System calibration.} Before each measurement session, the microphone channels were calibrated using a Brüel \& Kjær 4231 acoustic calibrator at a 94\,dB SPL reference level.
\item \textbf{Experimental configuration selection.} The room configuration (\texttt{RC}), loudspeaker configuration (\texttt{SC}), and microphone modality were selected according to the experimental design.
\item \textbf{Excitation playback.} A sweep-based excitation signal covering the audible range was reproduced through each active loudspeaker in the selected configuration.
\item \textbf{Multichannel recording.} All microphone channels were recorded simultaneously using the multichannel acquisition system.
\item \textbf{Impulse-response extraction.} The recorded signals were deconvolved with the inverse sweep in order to obtain the impulse responses corresponding to each source--receiver pair.
\item \textbf{Storage and indexing.} The resulting responses were stored in MATLAB format and associated with their experimental context through the metadata and indexing structure of the dataset.
\end{enumerate}

This workflow ensures that every impulse response stored in the database can be related to its measurement conditions, including room configuration, loudspeaker configuration, microphone modality, and spatial location.


\subsection{Dataset statistics}

The current release of UPV\_RIR\_DB includes 166 multichannel RIR files, corresponding to a total of 18{,}976 individual impulse responses distributed across rooms, modalities, and experimental configurations. Table~\ref{tab:dataset_statistics} summarizes the number of multichannel RIR measurements and the corresponding total number of individual impulse responses, organized by room and acquisition parameters.

\begin{table*}[!t]
\centering
\caption{UPV\_RIR\_DB dataset statistics. The table summarizes the multichannel RIR measurements and the corresponding total number of individual impulse responses, organized by room and acquisition parameters.}
\label{tab:dataset_statistics}
\begin{tabular}{lcccccc}
\toprule
Room & Mic. modality & RIR files & Sources & Channels & IR per file & Total IRs \\
\midrule
8G4C & AR & 10 & 8 & 32 & 256 & 2560 \\
4P13 & AR & 32 & 8 & 32 & 256 & 8192 \\
4P13 & DH & 32 & 8 & 4 & 32 & 1024 \\
4D26 & AR & 19 & 8 & 32 & 256 & 4864 \\
4D26 & DH & 19 & 8 & 4 & 32 & 608 \\
4D26 & BI & 54 & 8 & 4 & 32 & 1728 \\
\midrule
\textbf{Total} & --- & \textbf{166} & --- & --- & --- & \textbf{18,976} \\
\bottomrule
\end{tabular}
\end{table*}


\section{Database design and structure}
\label{sec:design}

UPV\_RIR\_DB was designed to guarantee traceability, reproducibility, and clarity in the interpretation of measured impulse responses. Its organization combines a deterministic directory hierarchy with structured descriptive resources used to document and index the experimental context associated with each measurement.

The dataset organization is based on three main principles:

\begin{itemize}
\item \textbf{Directory hierarchy}, reflecting room configuration, loudspeaker configuration, and microphone modality.
\item \textbf{Structured descriptive resources}, documenting metadata, hardware references, internal conventions, and associated acoustic descriptors.
\item \textbf{Centralized indexing}, allowing each stored RIR file to be linked to its experimental context.
\end{itemize}


\subsection{Hierarchy and conventions}

Each room dataset is stored inside a directory following the convention

\begin{verbatim}
      data/Room_<room_id>_mat/
\end{verbatim}    

where \texttt{<room\_id>} uniquely identifies the room within the corpus, e.g., \texttt{Teleco\_4P13}, \texttt{Teleco\_4D26}, or \texttt{iTEAM\_8G4C}. In this convention, the room identifier combines an institutional or site label with a compact location code separated by an underscore. For example, \texttt{Teleco\_4P13} identifies a room associated with the \textit{Escuela T\'{e}cnica Superior de Ingenieros de Telecomunicai\'{o}n} environment, while \texttt{4P13} acts as a compact code for the corresponding UPV building (4P)\footnote{The building numerology of the UPV campus is available at their website: \url{https://www.upv.es/plano/index-en.html}} and specific classroom (13). This criterion promotes compact, readable, and consistent identifiers across the corpus.

The dataset adopts a Cartesian coordinate convention to ensure that the spatial descriptors associated with zones, loudspeakers, and microphones are mutually consistent and reproducible. For rooms with a single main loudspeaker wall, the origin \((0,0,0)\) is defined at floor level, at the corner of the wall opposite the source wall. The \(x\)-axis extends from left to right when facing the loudspeaker wall, the \(y\)-axis extends from the origin wall toward the source wall, and the \(z\)-axis represents height from the floor.

Room \texttt{8G4C} constitutes a particular case because it includes two opposite loudspeaker placements, \texttt{SC1} and \texttt{SC2}, both oriented toward the central measurement region. In this room, a single room-fixed coordinate system is used for the complete dataset, with the origin located at the lower-left floor corner of the room layout. Under this convention, \texttt{SC2} is located near the lower wall of the room and \texttt{SC1} near the upper wall, while all zones, points, and source coordinates remain expressed within the same spatial reference. These coordinate conventions are stored in the \texttt{coordinate\_convention} field of the corresponding room metadata and provide the reference used to interpret all spatial coordinates in the dataset.

Within each room directory, measurements are organized according to two experimental configuration descriptors:

\begin{itemize}
\item \textbf{Room Configuration (RC):} defines the physical or acoustic condition of the room during the measurement session.
\item \textbf{Speaker Configuration (SC):} defines the loudspeaker system used in the experiment, including the number of active sources and their spatial arrangement.
\end{itemize}

Separating RC and SC allows distinguish between changes caused by the acoustic environment and changes caused by the excitation layout.

In addition to the main directory hierarchy used to store the RIRs, the dataset relies on several descriptive and auxiliary resources with different scopes. These include: \textit{(i)} a global master file with consolidated spatial information organized by room, \textit{(ii)} the structured room-level metadata in JSON and MAT formats, \textit{(iii)} local metadata associated with selected BI measurements at the RIR level, \textit{(iv)} hardware reference files used for labeling and traceability, and \textit{(v)} mapping resources used to maintain internal conventions, equivalences, and transformations associated with migration and normalization processes. The organization and role of these resources are described later in the metadata subsection.

The directory hierarchy is complemented by a systematic file-naming convention for the RIR measurements. File names inherit the logic of the directory structure and encode, in compact form, the most relevant experimental context of each acquisition. To promote standardization, automated parsing, and dataset scalability, the database adopts consistent labels, controlled uppercase usage, and homogeneous alphanumeric formatting whenever possible. This criterion facilitates indexing, search, filtering, and integration into reproducible processing pipelines, including automated analysis and machine-learning-based workflows.


\subsection{Directory structure}

The physical organization of the dataset follows the experimental hierarchy described above, as illustrated in Fig.~\ref{fig:dir_structure_en}. Within each room dataset, the directory structure reflects the room configuration (RC), loudspeaker configuration (SC), and microphone modality used during acquisition.

In this hierarchy:

\begin{itemize}
\item \texttt{RC<r>} identifies the room configuration used during the measurement.
\item \texttt{SC<s>} identifies the loudspeaker configuration.
\item \texttt{AR}, \texttt{DH}, and \texttt{BI} identify the microphone modality used for acquisition.
\item Multichannel RIR files are stored directly inside the directory corresponding to each microphone modality.
\item The \texttt{RT} subdirectory contains auxiliary MATLAB files with reverberation-related descriptors derived from those measurements.
\end{itemize}

In addition, each room dataset includes the files \texttt{metadata.json} and \texttt{metadata.mat}, stored at the top level of the room directory, where the complete semantic description of the measurements associated with that room is defined. In this way, the directory hierarchy provides the physical organization of the files, while the metadata define their experimental, spatial, and acoustic interpretation.

\begin{figure}[!t]
\centering
\begin{verbatim}
data/
  Room_<room_id>_mat/
    metadata.json
    metadata.mat
    RC<r>/
      SC<s>/
        AR/
          <RIR files>
          RT/
            <RT files>
        DH/
          <RIR files>
          RT/
            <RT files>
        BI/
          <RIR files>
          RT/
            <RT files>
\end{verbatim}
\caption{Directory structure of UPV\_RIR\_DB. Multichannel RIR files are stored directly inside each microphone-modality directory, while the \texttt{RT} subdirectory contains auxiliary reverberation descriptors derived from those measurements.}
\label{fig:dir_structure_en}
\end{figure}


\subsection{Metadata organization}

Each room dataset includes a room-level metadata description, stored in the files \texttt{metadata.json} and replicated in \texttt{metadata.mat}. These files contain the structured description of the experimental, spatial, and acoustic context associated with the impulse responses stored for that room. The metadata structure is organized into two top-level blocks: \texttt{general} and \texttt{room}.

The \texttt{general} block gathers descriptors that apply to the complete room dataset. In the current structure, it includes the fields \texttt{coordinate\_convention}, \texttt{process\_factors}, \texttt{signal}, \texttt{hardware}, \texttt{calibrations}, \texttt{analysis}, \texttt{configuration\_map}, and \texttt{analysis\_totals}. These fields describe, respectively, the coordinate-system conventions, processing factors, excitation-signal parameters, hardware references, calibration references, analysis information, configuration maps, and aggregated acoustic descriptors at room level. Taken together, this block concentrates dataset-wide information that does not depend on an individual measurement, but instead provides a common reference for interpretation, traceability, and internal dataset organization.

The \texttt{room} block contains the information specific to the measured environment and its internal organization. In the current structure, it includes the fields \texttt{id}, \texttt{dimensions}, \texttt{room\_configurations}, \texttt{speaker\_configurations}, \texttt{speaker\_sets}, \texttt{positions}, \texttt{trajectories}, and \texttt{index}. Together, these fields describe the room identity, its physical dimensions, the available experimental configurations, the spatial organization of the measurement campaign, and the linking layer between stored files and measurement context.

Within the \texttt{room} block, spatial descriptors are stored under the \texttt{positions} field, which includes three main elements: \texttt{positions.zones}, \texttt{positions.loudspeakers}, and \texttt{positions.microphones}. The first defines the catalog of measurement zones and their associated points; the second stores the spatial coordinates of the loudspeakers for each excitation configuration; and the third stores the three-dimensional coordinates of the microphone elements associated with each acquisition modality. This separation makes it possible to distinguish between the abstract spatial organization of the campaign and the effective physical geometry of the devices involved.

The \texttt{trajectories} field, also included in \texttt{room}, defines ordered sequences of zones used to represent spatial paths within the general measurement scheme.
In addition to the ordered zone sequence, trajectory entries may include qualitative descriptors such as \texttt{pattern}, \texttt{alignment}, \texttt{monotonicity}, and \texttt{axis\_reference}. These fields are used as metadata descriptors for classification, query, and analysis purposes, and do not modify the underlying spatial coordinates.

In turn, the \texttt{index} field acts as the linking layer between the stored RIR files and their complete experimental context. Each index entry may associate a file with its microphone modality, room configuration, loudspeaker configuration, zone, associated trajectories, and descriptors linked to that measurement. In this sense, although \texttt{index} belongs structurally to the \texttt{room} block, functionally it serves as the integrative mechanism between physical hierarchy, spatial descriptors, and stored measurements.

With this structure, the room-level metadata do not merely describe the dataset contents, but establish an explicit framework for connecting geometry, experimental configurations, hardware references, stored measurements, and associated descriptors within a uniform and traceable organization.


\subsection{Acoustic descriptors and reverberation metadata}

In addition to the geometric and experimental organization of the dataset, the database incorporates acoustic descriptors intended to characterize the reverberant behavior of the stored impulse responses. In the current structure, these descriptors are represented at two complementary levels: \textit{individual-measurement level} and \textit{room-aggregated level}.

At measurement level, the \texttt{index} field within the \texttt{room} block may associate each RIR file with its corresponding acoustic descriptors. In this way, an index entry not only links the file to its microphone modality, room configuration, loudspeaker configuration, zone, and associated trajectories, but also to the acoustic information derived from that specific measurement. This organization preserves traceability between each stored RIR and the acoustic parameters obtained from it.

At aggregated level, the \texttt{general.analysis\_totals} block stores global acoustic descriptors of direct interest for the room. These values provide a compact and directly interpretable representation of the reverberant behavior of the enclosure, while remaining consistent with the detailed results associated with individual measurements.

The adopted nomenclature distinguishes between ``interval-based quantities directly measured on the decay curve'' and ``reverberation estimators extrapolated to 60 dB''. The non-extrapolated interval-based quantities are stored using the following field names:

\noindent\hspace*{1em}\texttt{edt\_0\_10\_raw}: decay interval between \(0\) and \(-10\) dB;\\
\noindent\hspace*{1em}\texttt{t20\_5\_25\_raw}: decay interval between \(-5\) and \(-25\) dB;\\
\noindent\hspace*{1em}\texttt{t30\_5\_35\_raw}: decay interval between \(-5\) and \(-35\) dB.

From these quantities, the database defines the extrapolated estimators \texttt{edt}, \texttt{t20}, and \texttt{t30}, obtained by projecting those intervals to a theoretical 60 dB decay. The calculation is based on the integrated energy decay curve obtained through the classical Schroeder backward integration method and on the standardized EDT, T20, and T30 definitions commonly adopted in ISO 3382-1~\cite{Schroeder1965,ISO3382-1}. The relation between both representations is given by:

\begin{align}
\mathtt{edt} &= 6 \cdot \mathtt{edt\_0\_10\_raw} \\
\mathtt{t20} &= 3 \cdot \mathtt{t20\_5\_25\_raw} \\
\mathtt{t30} &= 2 \cdot \mathtt{t30\_5\_35\_raw}
\end{align}

The descriptor \texttt{rt60} is used as the final reverberation-time value reported by the database, typically prioritizing the most suitable available estimate according to the adopted processing criterion. In turn, \texttt{rt60\_real} is retained as an auxiliary direct estimate of the measured decay, and may be stored as \texttt{NaN} when dynamic range or signal-to-noise limitations do not allow a robust estimation.

With this organization, the dataset preserves both the detailed acoustic characterization of each individual measurement and an aggregated room-level representation, using an explicit and traceable nomenclature that facilitates interpretation, comparison across measurements, and the integration of these descriptors into later analysis workflows.

\section{Conclusions}
\label{sec:concl}

This paper presented UPV\_RIR\_DB, a structured database of measured room impulse responses designed to provide reproducible acoustic data with explicit spatial metadata and hierarchical organization. The dataset currently includes 18,976 impulse responses measured in three rooms of the Universitat Polit\`{e}cnica de Val\`{e}ncia under different room configurations, loudspeaker layouts, and microphone modalities.

The database design combines a directory hierarchy with structured metadata files describing acquisition parameters, spatial geometry, and calibration references. In particular, the metadata separates global acquisition descriptors from room-level spatial descriptors such as zones, loudspeaker coordinates, and microphone positions. A central index links each stored RIR file with its measurement context, allowing every impulse response to be traced to its experimental conditions.

This organization provides not only a dataset of measured impulse responses but also a structured framework for storing, analyzing, and reusing acoustic measurements in a reproducible manner.

Future work will focus on extending the dataset with additional rooms, expanding the catalog of measurement zones, and incorporating further acoustic descriptors derived from the recorded impulse responses.

\section*{Data availability}

The UPV\_RIR\_DB dataset is publicly available through the Zenodo research data repository. The current version of the database can be accessed at \url{https://doi.org/10.5281/zenodo.19051416}. The repository includes the RIR measurements, the associated metadata files (\texttt{metadata.json} and \texttt{metadata.mat}), and documentation describing the database structure. Future releases of the dataset will be distributed through the same Zenodo record to ensure long-term accessibility, reproducibility, and version traceability.


\appendices

\section{Extended metadata schema}
\label{app:metadata}

This appendix summarizes the structure of the metadata associated with each room dataset. The metadata is stored as \texttt{metadata.json} and mirrored in MATLAB format as \texttt{metadata.mat}. The scheme is organized in two main blocks: \texttt{general} and \texttt{room}. All names use ASCII characters and lowercase identifiers with underscores. SI units are used unless otherwise stated.

\subsection{General descriptors}

The \texttt{general} block contains descriptors that apply to the entire room dataset:

\begin{itemize}[leftmargin=*,nosep]
  \item \texttt{coordinate\_convention}: definition of the coordinate system used in the dataset.
  \item \texttt{process\_factors}: parameters related to the acquisition and processing pipeline.
  \item \texttt{signal}: excitation signal parameters used for impulse response measurements.
  \item \texttt{hardware}: description of microphones, loudspeakers and acquisition devices.
  \item \texttt{calibrations}: references to microphone calibration files and associated metadata.
  \item \texttt{analysis}: configuration parameters used during acoustic analysis.
  \item \texttt{configuration\_map}: mapping between room configurations (RC) and speaker configurations (SC).
  \item \texttt{analysis\_totals}: global acoustic descriptors summarizing the room measurements.
\end{itemize}

\subsection{Room descriptors}

The \texttt{room} block describes the physical environment and spatial organization of the measurements.

\begin{itemize}[leftmargin=*,nosep]
  \item \texttt{id}: canonical room identifier.
  \item \texttt{dimensions}: room dimensions expressed in meters.
  \item \texttt{room\_configurations}: available room configurations (RC).
  \item \texttt{speaker\_configurations}: loudspeaker configurations (SC).
  \item \texttt{speaker\_sets}: definition of the loudspeaker sets used during measurements.
\end{itemize}

\subsection{Spatial descriptors}

Spatial descriptors are stored under the \texttt{positions} field.

\begin{itemize}[leftmargin=*,nosep]

\item \texttt{positions.zones}: catalog of spatial zones used to organize the measurement grid. Each zone contains a center coordinate and a set of measurement points with explicit coordinates defined within the zone.

\item \texttt{positions.loudspeakers}: coordinates of the loudspeakers for each speaker configuration.

\item \texttt{positions.microphones}: three-dimensional coordinates of the microphone capsules for each microphone modality.

\end{itemize}

Zones describe the abstract spatial grid used during the measurement campaign, whereas microphone coordinates represent the physical receiver geometry.

\subsection{Trajectories}

The \texttt{trajectories} field defines ordered sequences of zones representing measurement paths. Trajectories do not introduce new spatial coordinates but reference zones already defined in the dataset.

\begin{figure*}[t]
\centering
\caption{Minimal JSON snippet illustrating the metadata structure.}
\label{lst:json-min}
\lstset{
  language=json,
  basicstyle=\ttfamily\tiny,
  showstringspaces=false,
  breaklines=true,
  breakatwhitespace=false,
  columns=fullflexible,
  frame=single,
  xleftmargin=0pt,
  xrightmargin=0pt,
  aboveskip=0pt,
  belowskip=0pt
}
\begin{minipage}[t]{0.485\textwidth}
\vspace{0pt}
\begin{lstlisting}
{
  "general": {
    "coordinate_convention": {
      "origin_xy": "opposite_to_sources_wall",
      "x_axis": "left_to_right (...)",
      "y_axis": "wall_towards_sources"
    },

    "process_factors": {
      "fs": 44100,
      "amplifier_gain": 5,
      ...
    },

    "signal": {
      "type": "ESM",
      "f_ini": 20,
      "f_end": 22050,
      "duration": 4,
      ...
    },

    "hardware": {
      "audio_interfaces": ["ifc_001"],
      "loudspeakers": ["spk_001"],
      "microphones": {
        "AR": ["mic_AR_001"],
        "BI": ["mic_BI_001"],
        "DH": ["mic_DH_001"]
      }
    },

    "calibrations": {
      "AR": {
        "micID": "16_MIC_Bruel_array STUDIO-CAPTURE",
        "ref_SPL": 94,
        "insertion_loss_dB": 1.7,
        "freq": 1000,
        "date": "2025_4_14",
        "factors": [254.94, 249.82, ...]
      },
      "BI": {
        "micID": "...",
        ...
      },
      "DH": {
        "micID": "...",
        ...
      }
    },

    "analysis": {
      "metadata_version": "v1",
      "creation_date": "2025-11-10",
      "metrics_version": "v1.0_rt_broadband",
      "methods": "Broadband EDT/T20/T30/RT60 ...",
      "notes": "Global RT based on ref_mic_type=AR."
    },

    "configuration_map": {
      "St": "RC4",
      "Sv": "RC3",
      ...
    },

    "analysis_totals": {
      "AR": {
        "edt": 0.68,
        "t20": 0.89,
        "t30": 0.81,
        "rt60": 0.81,
        "edt_0_10_raw": 0.11,
        "t20_5_25_raw": 0.29,
        "t30_5_35_raw": 0.40,
        "rt60_real": 0.34
      }
    }
  },
\end{lstlisting}
\end{minipage}
\hfill
\begin{minipage}[t]{0.485\textwidth}
\vspace{0pt}
\begin{lstlisting}
  "room": {
    "id": "Room_Teleco_4D26_mat",

    "dimensions": {
      "width": 5.85,
      "length": 11.73,
      "height": 2.98
    },

    "room_configurations": ["RC1"],
    "speaker_configurations": ["SC1"],

    "speaker_sets": {
      "SC1": ["spk_001"]
    },

    "positions": {
      "zones": {
        "Z01": {
          "center_xy": [5.4075, 3.6625],
          "points_xy": {
            "P01": [5.39, 3.645],
            "P02": [5.425, 3.68]
          }
        },
        ...
      },

      "loudspeakers": {
        "SC1": {
          "src1": [5.44, 4.9, 1.65],
          "src2": [5.6275, 4.9, 1.65],
          ...
        }
      },

      "microphones": {
        "AR": {
          "mic_AR_001": {
            "measurement_count": 32,
            "coordinates": {
              "Z01": {
                "rcv1": [5.495, 3.75, 1.65],
                "rcv2": [5.425, 3.75, 1.65],
                ...
              }
            }
          }
        },
        "BI": { ... },
        "DH": { ... }
      }
    },

    "trajectories": {
      "T01": {
        "zones": ["Z15","Z12","Z05","Z09","Z02"],
        "pattern": "diagonal",
        ...
      }
    },

    "index": [
      {
        "rir_file": "AR_RC1_SC1_Z01_RIR.mat",
        "mic_type": "AR",
        "room_config": "RC1",
        "speaker_config": "SC1",
        "zone": "Z01",
        "trajectories": ["T01"],
        "acoustics": {
          "edt": 0.62,
          "t20": 0.88,
          "t30": 0.81,
          "rt60": 0.81,
          "rt60_real": 0.34,
          ...
        }
      }
    ]
  }
}
\end{lstlisting}
\end{minipage}
\end{figure*}

\paragraph{Example}

A trajectory is defined as an ordered list of zones together with a set of descriptive attributes. For instance:

\begin{verbatim}
"T01": {
  "zones": ["Z15","Z12","Z05","Z09","Z02"],
  "pattern": "diagonal",
  "alignment": "undefined",
  "monotonicity": "increasing",
  "axis_reference": "xy",
  "notes": ""
}
\end{verbatim}

The \texttt{zones} field defines the ordered path, while the remaining fields provide additional descriptors of the trajectory geometry and interpretation. These attributes are used for analysis purposes and do not modify the underlying zone definitions.

\subsection{Central index}

The \texttt{index} field provides the mapping between stored RIR files and their experimental context. Each entry associates a file with:

\begin{itemize}[leftmargin=*,nosep]
  \item room identifier
  \item room configuration (RC)
  \item speaker configuration (SC)
  \item microphone modality
  \item zone identifier
  \item trajectory identifier when applicable
  \item an \texttt{acoustics} block containing acoustic descriptors derived from the impulse response (e.g., EDT, T20, T30, RT60 estimates)
\end{itemize}

Fig.~\ref{lst:json-min} illustrates a minimal JSON example of the metadata structure.

The index acts as the authoritative mapping of the dataset and enables structured queries across measurements. 
\section{Measurement Hardware}
\label{app:hardware}

This appendix summarizes the hardware used during the measurement campaigns. All devices are registered in the dataset under a unified hardware reference system, allowing consistent identification across rooms and measurement configurations.

\subsection{Audio Interfaces}

The recordings were performed using multichannel audio interfaces. The system is based on the following device:

\begin{itemize}
\item \textbf{ROLAND STUDIO-CAPTURE 16 x 8} (\texttt{ifc\_001}): 16 input channels and 8 output channels using ASIO drivers. This interface was used as the main acquisition unit, and multiple units were synchronized when higher channel counts were required.
\end{itemize}

\subsection{Loudspeakers}

The excitation system is based on the following loudspeaker model:

\begin{itemize}
\item \textbf{JBL LSR305} (\texttt{spk\_001}): active studio monitor used as sound source. The measurement setup employs an 8-source configuration, with spatial positions defined in the metadata of each room.
\end{itemize}

\subsection{Microphones}

Three microphone modalities are included in the dataset, each corresponding to a specific acquisition purpose:

\begin{itemize}
\item \textbf{Array microphone (AR)} (\texttt{mic\_AR\_001}): Brüel \& Kjær Type 4951 measurement microphones. Condenser-based sensors designed for high-precision acoustic measurements, used as the main multichannel receiver.

\item \textbf{Dummy head microphone (DH)} (\texttt{mic\_DH\_001}): Neumann KU100 binaural head. A condenser-based dummy head system used for perceptually oriented binaural recordings.

\item \textbf{Binaural headset (BI)} (\texttt{mic\_BI\_001}): Brüel \& Kjær Type 4101-A binaural microphone set. Miniature condenser microphones designed for in-ear recordings, used in trajectory-based measurements.
\end{itemize}

All hardware elements are referenced in the dataset metadata through unique identifiers, enabling consistent linkage between measurement configurations, acquisition devices, and recorded data.

\balance
\bibliographystyle{IEEEtran}
\bibliography{upv_rir_db}

\end{document}